\documentclass{article}
\usepackage{spconf,amsmath,graphicx}
\usepackage[T1]{fontenc}
\usepackage{pifont}
\usepackage{tabularx}
\usepackage{multirow}
\usepackage{amsmath}
\usepackage{amsfonts}
\usepackage{amssymb}
\usepackage{wrapfig}
\usepackage{tikz}
\usepackage{comment}
\usepackage{amsmath,amssymb}
\usepackage{color}
\usepackage{multirow}
\usepackage{kotex}
\usepackage{adjustbox}
\usepackage{float}
\usepackage{booktabs}
\usepackage{colortbl}
\usepackage{graphicx,verbatim}
\usepackage{url}
\usepackage{enumitem}
\setlist{nosep, leftmargin=14pt}

\usepackage{mwe}

\title{Rethinking Glaucoma Calibration: Voting-Based Binocular and Metadata Integration}

\name{ 
Taejin Jeong\textsuperscript{1, 2} \quad
Joohyeok Kim\textsuperscript{1} \quad
Jaehoon Joo\textsuperscript{3} \quad
Seong Jae Hwang\textsuperscript{1}\textsuperscript{\(\star\)}\thanks{\textsuperscript{\(\star\)} Corresponding author.}
}

\address{
$^{1}$Yonsei University, Seoul, Republic of Korea\\
$^{2}$Mediwhale, Seoul, Republic of Korea\\
$^{3}$LG Electronics, Seoul, Republic of Korea\\
}

\begin{document}

\maketitle

\begin{abstract}
\noindent{Glaucoma is a major cause of irreversible blindness, with significant diagnostic subjectivity.} 
This inherent uncertainty, combined with the overconfidence of models optimized solely for accuracy can lead to fatal issues such as overdiagnosis or missing critical diseases. 
To ensure clinical trust, \textit{model calibration} is essential for reliable predictions, yet study in this field remains limited. 
Existing calibration study have overlooked glaucoma's systemic associations and high diagnostic subjectivity. 
To overcome these limitations, we propose V-ViT (Voting-based ViT), a framework that enhances calibration by integrating a patient's binocular information and metadata. 
Furthermore, to mitigate diagnostic subjectivity, V-ViT utilizes an \textit{iterative dropout-based Voting System} to maximize calibration performance. 
The proposed framework achieved state-of-the-art performance across all metrics, including the primary calibration metrics. 
Our results demonstrate that V-ViT effectively resolves the issue of overconfidence in predictions in glaucoma diagnosis, providing highly reliable predictions for clinical use. 
Our source code is available at \url{https://github.com/starforTJ/V-ViT}.
\end{abstract}

\section{Introduction}
\label{sec:intro}
Glaucoma currently affects 80 million individuals globally and is a leading cause of blindness, with incidence rates on the rise~\cite{glaucoma_percent}. 
Owing to its asymptomatic nature, glaucoma often remains undetected in its incipient stages, leading to substantial vision impairment and escalating treatment expenditures for advanced cases. 
Therefore, early and accurate diagnosis is paramount for preventing irreversible vision loss and mitigating financial burdens~\cite{earlydetection}. 
However, due to the high cost of optical coherence tomography (OCT), routine screening relies on fundus photography and visual field testing, which are subject to high diagnostic subjectivity and significant inter-observer variability~\cite{kappa, inter1}. 
This convergence of high clinical risk and high subjectivity emphasizes the need for \textit{model calibration} in AI-driven diagnostic tools, alongside high accuracy.

Model calibration is the fundamental process of adjusting a AI model's predicted probabilities to align with the empirical likelihood of an event's occurrence. 
This ensures that in a population of samples for which the model predicts a $90\%$ probability of glaucoma, the actual proportion of positive cases is indeed $90\%$. 
While recent AI models achieve high classification accuracy due to their deep and complex structures, their model calibration performance tends to degrade, often resulting in overconfidence. 
In the medical domain, this overconfidence poses serious problems. 
Overconfidence in a positive diagnosis can induce clinicians to rush into unnecessary, invasive, and high-cost additional tests or immediate aggressive treatments. 
Conversely, overconfidence in the absence of the disease risks the clinician overlooking a critical early diagnosis. 
Therefore, regardless of how high the model's accuracy is, if its predicted probability values do not accurately reflect the true risk, the quality of clinical decision-making will inevitably decline. 
For this reason, model calibration is a fundamental element in medical AI models, just as important as accuracy. 
Although an initial study on glaucoma calibration~\cite{glaucoma_calibration} has emerged, it shows limitations by failing to comprehensively address the disease's unique pathophysiological characteristics. 
Accordingly, this study aims to resolve this calibration challenge by leveraging the fact that glaucoma has a close etiological relationship with systemic diseases and by accounting for the inherent inter-observer variability in the clinical diagnostic process.

\begin{figure*}[t]
    \centering
    \includegraphics[width=\textwidth]{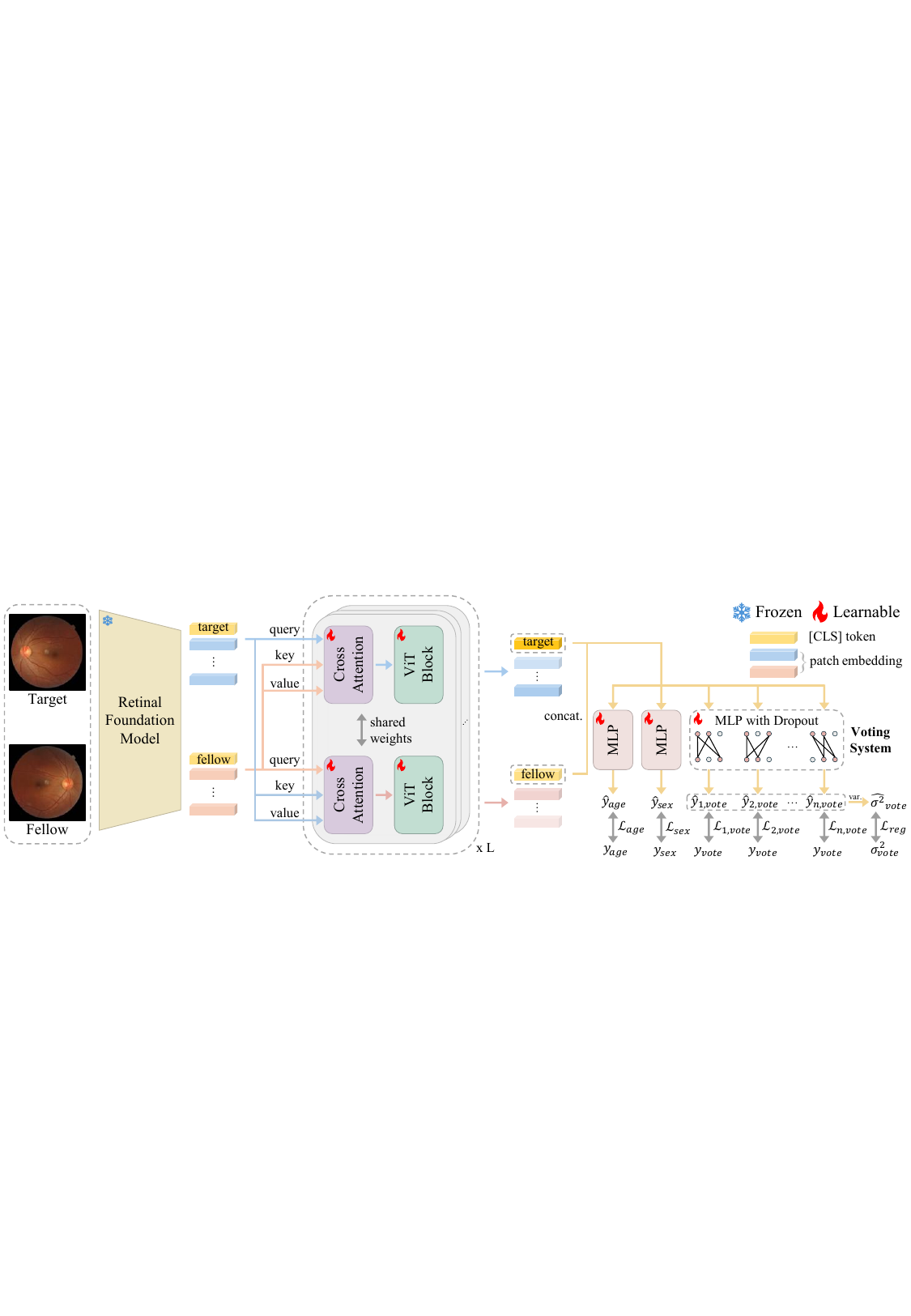}
    \caption{Overview of V-ViT. Through a total of $L$ Cross-Attention blocks, V-ViT learns to incorporate the fellow image and metadata for glaucoma diagnosis.}
    \label{fig:Model pipeline}
\end{figure*}

To achieve superior calibration performance, we address glaucoma's unique challenges by capitalizing on two main attributes. 
First, we utilize the fact that glaucoma is closely associated with the fellow eye and systemic diseases~\cite{bothcorr, complexity1, glaucoma_age, glaucoma_sex}. 
We resolve the calibration problem by fusing binocular data through a Cross-Attention block. 
Furthermore, by training the model as a multi-task model to simultaneously predict age and sex, we induce the system to implicitly learn crucial systemic factors, thereby enhancing overall calibration performance. 
Second, we construct a novel dataset to mitigate the high subjectivity caused by inter-observer variability. 
This dataset uses the average of binary classifications (0 and 1) made by multiple trusted ophthalmologists as the label. 
These averaged label values accurately reflect the distribution of actual physician judgments, satisfying the intent of model calibration while simultaneously yielding an indirect label smoothing effect~\cite{labelsmooth}. 
Building on this, we propose the V-ViT (Voting-based ViT) framework, which is built upon the ViT~\cite{ViT} architecture and incorporates a Monte Carlo (MC) dropout-based Voting System to solve the calibration problem in diagnostic tasks with high subjectivity.
 
\noindent
\textbf{Our contributions:}
\textbf{(1)} We address the calibration problem by leveraging the characteristics of glaucoma through binocular data and metadata; 
\textbf{(2)} We propose a new approach to address calibration in highly subjective diagnoses and introduce a Voting System as a solution, along with a new framework, V-ViT;
\textbf{(3)} The proposed V-ViT model achieves State-of-the-Art performance across all evaluation metrics.

\section{Methodology}
\label{sec:methods}
Fig.~\ref{fig:Model pipeline} illustrates the overall architecture of our proposed framework, V-ViT. Our framework learns a disease-specific representation by integrating binocular information and metadata, and subsequently performs a reliable diagnosis via an iterative dropout-based Voting System.

\subsection{Glaucoma-specific Representation Learning}
In this section, we describe our approach to learning glaucoma-specific representation, which integrates two crucial sources of information: the glaucoma's bilateral characteristics and patient-specific metadata.
To account for the bilateral nature of glaucoma, V-ViT utilizes images from both the target and fellow eyes. Initial embeddings are extracted from each image via a pretrained retinal foundation model. 
Subsequently, Cross-Attention is employed to fuse their complementary information. Cross-Attention modules, in which the embeddings from each eye act as queries to each other's information, enable the model to capture features that are difficult to discern from a single-eye alone.

Furthermore, V-ViT employs a multi-task learning approach to incorporate patient-specific factors such as age and sex. The final representation $z$ is formed by concatenating the [CLS] tokens from the target and fellow eyes after the final ViT block. This representation is then fed into three separate MLP heads to simultaneously predict glaucoma, age, and sex. 
This multi-task scheme acts as a regularization with clinical inductive bias. By compelling the model to simultaneously predict significant patient factors like age, it is forced to learn the underlying clinical trends (e.g., age as a primary risk factor) associated with the disease. This process prevents the model from overfitting to the primary task and encourages it to learn a more robust and generalized representation that not only encapsulates patient characteristics but also implicitly improves prediction calibration.
While we utilize easily obtainable age and sex information, the framework is readily extendable to additional metadata, such as genetic data.

\newcolumntype{C}[1]{>{\hsize=#1\hsize\centering\arraybackslash}X}

\newcolumntype{L}[1]{>{\hsize=#1\hsize\raggedright\arraybackslash}X} 

\begin{table*}[t]
    \caption{Quantitative evaluation of V-ViT performance compared to Vijayan et al. and various baseline models, including CNN, ViT, and Linear Probing. Results from the ablation study are also included. Bolded values mark the best performance. (B: Binocular data, V: Voting System, and M: Metadata, respectively.)}
    \label{tab:Quantitative}
    \centering
    \footnotesize
    \setlength{\tabcolsep}{6pt}
    \begin{tabularx}{\linewidth}{@{}L{1.45} | C{0.25} C{0.25} C{0.25}|C{1.3} C{1.3} C{1.3} C{1.3} C{1.3} C{1.3}}
        \specialrule{0.5pt}{0pt}{2pt}
        Model & B & V & M & Recall $\uparrow$ & F1 $\uparrow$ & Brier $\downarrow$ & AUROC $\uparrow$ & ECE $\downarrow$ & ACC $\uparrow$ \\ 
        \specialrule{0.5pt}{1pt}{1pt}
        ResNet50 &  &  &  & 0.456 &  0.521 &  0.103 & 0.859  & 0.024  & 0.852   \\
        ViT-Tiny &  &  &  &  0.426 &  0.532 &  0.098 & 0.870 & 0.026 & 0.867 \\
        Linear Probing &  &  &  & 0.176 &  0.273 &  0.127 & 0.758 & 0.037 & 0.833 \\
        \specialrule{0.5pt}{1pt}{1pt}
        \multirow{1}{*}{Vijayan et al.~\cite{glaucoma_calibration}} &  &  &  &  0.412 &  0.487 &  0.110 & 0.840 & 0.048 & 0.846 \\ 
        \specialrule{0.5pt}{0pt}{0pt}
        \multirow{4}{*}{\textbf{V-ViT (Ours)}} 
        &  &  &  &  0.368 &  0.500 &  0.093 & 0.884 & 0.031 & \textbf{0.870} \\
        & \ding{51} &  &  &  0.397 &  0.519 &  0.090 & 0.893 & 0.026 & \textbf{0.870} \\
        & \ding{51} & \ding{51} &  &  0.412 &  0.519 &  0.092 & 0.886 & \textbf{0.021} & 0.865 \\
        & \cellcolor{gray!30}\ding{51} & \cellcolor{gray!30}\ding{51} & \cellcolor{gray!30}\ding{51} & \cellcolor{gray!30} \textbf{0.515} &  \cellcolor{gray!30}\textbf{0.583} &  \cellcolor{gray!30}\textbf{0.088} & \cellcolor{gray!30}\textbf{0.896} & \cellcolor{gray!30}\textbf{0.021} & \cellcolor{gray!30}\textbf{0.870}\\ 
        \specialrule{0.5pt}{0pt}{0pt}
    \end{tabularx}
\end{table*}

\subsection{Reliable Diagnosis via Voting System}
Building upon the robust representation generated in the previous section, this section details the `Voting System' through which V-ViT derives a reliable final diagnosis. The concatenated [CLS] token $z$ is fed into the MLP head for glaucoma diagnosis, which incorporates dropout layers. For a single input, the Voting System generates $n$ predictions by performing $n$ forward passes through the single glaucoma MLP head. 

Inspired by concepts from Multi-sample dropout~\cite{dropout} and Monte Carlo dropout~\cite{mcdrop}, our Voting System keeps dropout enabled during both training and inference to ensure variability among the prediction. It is worth noting that while the retinal foundation model and ViT blocks also contain dropout layers, we only apply the iterative forward passes to the glaucoma diagnosis MLP head. This design choice avoids the significant computational overhead associated with performing $n$ full forward passes through the entire backbone. This approach enhances the model's reliability and calibration performance by moving beyond a single point estimate. Denoting the $i$-th prediction as $\hat{y}_{i, vote}$, the model is trained on the average loss across all $n$ predictions. 
\[
    \mathcal{L}_{i,vote} = \Vert y_{vote} - \hat{y}_{i, vote}\Vert^2_2, \; 
	\mathcal{L}_{vote} = \frac{1}{n} \sum_{i=1}^{n} \mathcal{L}_{i, vote}.
\]
We employ the MSE loss, as the ground truth labels, $y_{vote}$ in our dataset are continuous values derived from the averaged assessments of multiple ophthalmologists.

\subsection{Overall Training Loss}
The training loss is defined as follows:
\[
	\mathcal{L} = \mathcal{L}_{vote} + \lambda_{age}\mathcal{L}_{age} + \lambda_{sex}\mathcal{L}_{sex} + \lambda_{reg}\mathcal{L}_{reg} + \lambda_{wd} \Vert \theta \Vert^2.
\]
$L_{age}$ and $L_{sex}$ are the loss terms for the auxiliary tasks of predicting the patient's age and sex, respectively. We apply the MSE loss for the age regression and the Cross-Entropy loss for the sex classification. 
$\mathcal{L}_{reg}$ is a regularization term that adjusts the variance of the V-ViT outputs to match the variance of assessments from multiple ophthalmologists, denoted as $\sigma^2_{vote}$ in Fig.~\ref{fig:Model pipeline}. 
This loss term is applied only for data samples assessed by three or more ophthalmologists to ensure the reliability of the target variance. 
It is crucial to distinguish this target variance from the arbitrary subjectivity of a single diagnosis we aim to resolve. The inter-observer variance among experts does not represent random error, but rather quantifies the inherent clinical difficulty and ambiguity of a case. This regularization term, therefore, directly calibrates the model's predictive uncertainty. By encouraging the variance of the model's predictions to match this meaningful variance of human experts, our model learns not only to be accurate but also to express a clinically meaningful degree of confidence in its predictions. 
$\lambda_{wd} \Vert \theta \Vert^2$ is introduced to impose a prior on $\theta$ in the context of Bayesian deep learning, where $\theta$ represents the parameters of V-ViT. $\lambda_{wd}$ is the corresponding weight decay coefficient.

\section{Experiments}
\label{sec:experiments}
\begin{figure}[t!]
    \centering
    \includegraphics[width=\columnwidth]{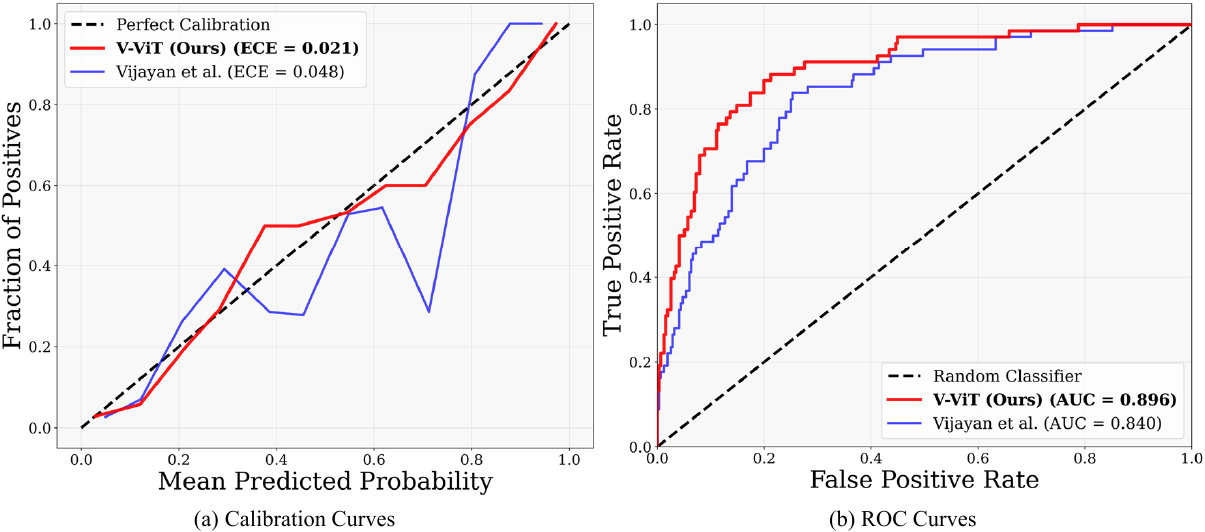}
    \caption{(a) The calibration curves comparing the model proposed by Vijayan et al. with V-ViT. (b) The ROC curves comparing the model proposed by Vijayan et al. with V-ViT.}
    \label{fig:Calibraion curve}
\end{figure}

\subsection{Experimental Setup}
We used data collected from GreenCross institution, where each sample was labeled by multiple ophthalmologists (an average of 3). 
After preprocessing, we obtained 3,830 samples in total. 
Among these, 3,446 samples were used for the training and validation dataset, while 384 samples were allocated to the test dataset. 
The $\text{PAPILA}$ dataset~\cite{papila} was employed for an independent external experiment. 
To focus on a binary classification task between \texttt{healthy} and \texttt{glaucoma}, we excluded samples from the \texttt{suspect} class. 
This filtering process yielded a curated dataset of 420 samples. 
Among these, 378 samples were used for the training and validation dataset, while 42 samples were allocated to the test dataset.

All experiments ran on a single NVIDIA RTX A6000 Ada GPU. 
Our V-ViT employed RETFound~\cite{retfound_mae} as the retinal foundation model, followed by $L=3$ Cross-Attention blocks. For the Voting System, we set the number of predictions $n=16$ and a dropout rate of 0.3.

\subsection{Quantitative Results}
Table~\ref{tab:Quantitative} compares the performance of V-ViT with that of the previous study using evaluation metrics commonly employed in medical calibration tasks. 
We performed a direct comparison against the work by Vijayan et al.~\cite{glaucoma_calibration}, which is the only preceding study focused on model calibration for glaucoma diagnosis. 
Furthermore, we compared our performance against widely-adopted models commonly used in medical image classification, along with the linear probing approach. 
Specifically, for linear probing, we utilized the embeddings from RETFound, the foundation model we employed.

We used evaluation metrics including Recall, F1, Brier, AUROC, Expected Calibration Error (ECE), and Accuracy (ACC). 
Among these, ECE and Brier are representative metrics of model calibration.
We achieved state-of-the-art results on all metrics. 
Low ECE scores and high Recall rates validate that our methodology is both safer and more accurate. 
Fig.~\ref{fig:Calibraion curve}(a) and Fig.~\ref{fig:Calibraion curve}(b) present the calibration curves and ROC curves comparing the model proposed by Vijayan et al. with V-ViT. 
As shown in Fig.~\ref{fig:Calibraion curve}(a), the calibration curve for V-ViT closely approximates the $y=x$ line, indicating it is well-calibrated. Furthermore, the ROC curve analysis in Fig.~\ref{fig:Calibraion curve}(b) confirms that V-ViT also achieves superior discriminative performance compared to the competing model.

We conducted a systematic ablation study to verify the performance contribution of our proposed methodology and determine the optimal $\text{V-ViT}$ architecture. 
Table~\ref{tab:Quantitative} illustrates the sequential performance gain achieved by integrating the binocular data, the Voting System, and metadata. 
The results confirm the synergistic and effective contribution of each component to the final performance.

\begin{table}[t]
    \caption{Quantitative Evaluation of V-ViT Performance and the model by Vijayan et al. on the PAPILA Dataset}
    \label{tab:Generalization}
    \centering
    \small
    \footnotesize
    \setlength{\tabcolsep}{6pt}
    \resizebox{\columnwidth}{!}{
        \begin{tabular}{l|cccccc}
            \specialrule{0.5pt}{0pt}{2pt}
            Model & Recall $\uparrow$ & F1 $\uparrow$ & Brier $\downarrow$ & AUROC $\uparrow$ & ECE $\downarrow$ & ACC $\uparrow$ \\ 
            \specialrule{0.5pt}{1pt}{1pt}
            \multirow{1}{*}{Vijayan et al.~\cite{glaucoma_calibration}} & 0.625 & 0.625 & 0.126 & 0.824 & 0.114 & 0.857 \\ 
            \cellcolor{gray!30}\multirow{1}{*}{\textbf{V-ViT (Ours)}} 
            & \cellcolor{gray!30}\textbf{1.000}  & \cellcolor{gray!30}\textbf{0.800} & \cellcolor{gray!30}\textbf{0.056} & \cellcolor{gray!30}\textbf{0.989} & \cellcolor{gray!30}\textbf{0.109} & \cellcolor{gray!30}\textbf{0.905} \\ 
            \specialrule{0.5pt}{0pt}{0pt}
        \end{tabular}
    }
\end{table}

\subsection{Generalization}
To demonstrate the robustness and generalization capability of V-ViT, we extended our evaluation beyond our custom dataset to an external, publicly available cohort. We selected the PAPILA dataset, an open-access resource that includes labels for binocular data, age, and sex, aligning perfectly with our approach. 
Table~\ref{tab:Generalization} presents the quantitative results comparing V-ViT against the baseline, Vijayan et al., on this external dataset. 
V-ViT consistently demonstrated superior performance across all metrics, and this cross-dataset validation decisively proved the efficacy and transportability of our proposed methodology in diverse clinical data environments.

\begin{figure}[t!]
    \centering
    \includegraphics[width=\columnwidth]{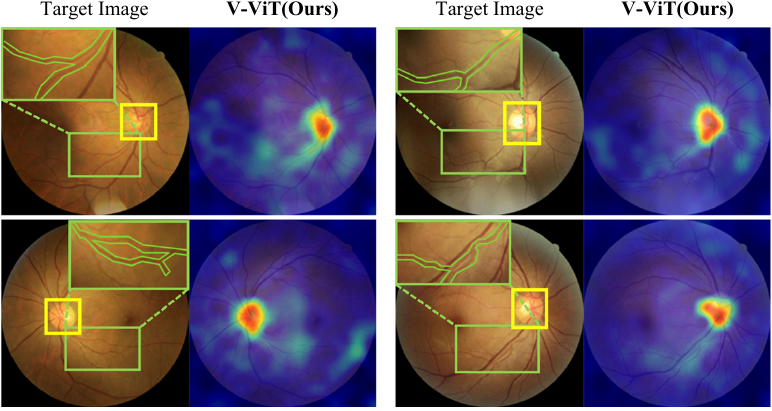}
    \caption{Attention map for the target image. The yellow boxes and green regions on the target image indicate the areas that should be focused on for glaucoma diagnosis.}
    \label{fig:attention map}
\end{figure}

\subsection{Model Interpretability}

Fig.~\ref{fig:attention map} presents the attention map generated using the final [CLS] token. 
This visualization clearly demonstrates the specific regions the model attends to during the final diagnostic decision-making process. 
V-ViT was confirmed to attend to clinically significant regions, typically observed by ophthalmologists, with a more precise and concentrated pattern. 

The yellow boxes indicate the optic disc and optic cup, which are the critical elements for glaucoma diagnosis in fundus images. Ophthalmologists primarily diagnose suspected glaucoma based on the ratio of these two structures ($\text{Cup/Disc ratio}$), and these defined regions provide the clinical basis for diagnosis. 
Furthermore the model focused on areas of optic nerve damage represented by the green regions. 
These areas are utilized for confirming glaucoma findings~\cite{wedge}, specifically showing attention to the wedge-shaped defect region associated with visual field loss.

This localization of attention signifies that the model accurately captures pathologically meaningful features beyond merely learning surrounding background patterns. 
This visual evidence serves as a key basis for enhancing the reliability and calibration of the model’s diagnostic predictions. Therefore, it suggests the potential for the model to serve as a powerful assistive tool that guides ophthalmologists to focus on key diagnostic areas during the diagnostic process, while simultaneously offering improved interpretability to increase confidence in its clinical adoption.

\section{Conclusion}
\label{sec:conclusion}
This study emphasizes the importance of calibration in glaucoma diagnosis. 
By integrating binocular information and metadata into our V-ViT, we effectively capture the bilateral and multi-faceted nature of the disease. 
Additionally, we introduce an MC dropout-based Voting System to mitigate overconfidence, thereby enhancing model reliability. 
Our model demonstrates superior performance on multiple metrics, reaffirming the significance of calibration in high-risk, highly subjective diseases like glaucoma. 
Moreover, this approach can be readily extended in clinical settings by incorporating additional metadata, offering a promising new pathway for developing reliable diagnostic models.
\clearpage

\section{Compliance with ethical standards}
\label{sec:ethics}

This study was performed in line with the principles of the Declaration of Helsinki. Approval was granted by the Ethics Committee of Public Institutional Review Board (PIRB) (Date: 2025-04-16 to 2029-12-31 / No.: P01-202409-01-004).

Additionally, this research study was conducted retrospectively using human subject data made available in open access by PAPILA. Ethical approval was not required as confirmed by the license attached with the open access data.

\section{Acknowledgments}
\label{sec:acknowledgments}

This work was supported in part by the IITP RS-2024-00457882 (AI Research Hub Project), IITP 2020-II201361, NRF RS-2024-00345806, NRF RS-2023-00219019, and NRF RS-2023-002620.

\bibliographystyle{IEEEbib}
\bibliography{ISBI2026}

\end{document}